# Unit testing, model validation, and biological simulation

**Gopal P. Sarma**[*1,6], **Travis W. Jacobs**[*2,6], **Mark D. Watts**[3,6], **S. Vahid Ghayoomie**[4,6], **Stephen D. Larson**[‡†6], **and Richard C. Gerkin**[‡5,6]

[1]**School of Medicine, Emory University, Atlanta, GA, 30307, USA (gopal.sarma@emory.edu)**
[2]**Department of Bioengineering, Imperial College London, London, UK (trav221@gmail.com)**
[3]**The University of Texas at Austin, Austin, TX, 78712, USA (wattsmark2015@gmail.com)**
[4]**Laboratory of Systems Biology and Bioinformatics, University of Tehran, Tehran, Iran (vahidghayoomi@gmail.com)**
[5]**School of Life Sciences, Arizona State University, Tempe, AZ, 85287, USA (rgerkin@asu.edu)**
[6]**OpenWorm Foundation, 1842 Commonwealth Ave, Unit 10, Boston, MA 02135 (stephen@openworm.org)**

**Abstract** The growth of the software industry has gone hand in hand with the development of tools and cultural practices for ensuring the reliability of complex pieces of software. These tools and practices are now acknowledged to be essential to the management of modern software. As computational models and methods have become increasingly common in the biological sciences, it is important to examine how these practices can accelerate biological software development and improve research quality. In this article, we give a focused case study of our experience with the practices of unit testing and test-driven development in *OpenWorm*, an open-science project aimed at modeling *Caenorhabditis elegans*. We identify and discuss the challenges of incorporating test-driven development into a heterogeneous, data-driven project, as well as the role of *model validation tests*, a category of tests unique to software which expresses scientific models.

[*]Joint first authors
[†]Corresponding author
[‡]Joint senior authors

# Introduction

Software plays an increasingly prominent role in the biological sciences. This growth has been driven by an explosion in the availability of data and the parallel development of software to store, share, and analyze this data. In addition, simulations have also become a common tool in both fundamental and applied research [1, 2]. Simulation management (initialization, execution, and output handling) relies entirely on software.

Software used for collaborative biological research has an additional level of complexity (beyond that shared by other widely-used software) stemming from the need to incorporate and interact with the *results* of scientific research, in the form of large data sets or dynamical models. This added level of complexity suggests that technical tools and cultural practices for ensuring software reliability are of particular importance in the biological sciences [3].

In this article, we discuss our experience in applying a number of basic practices of industrial software engineering—broadly known as *unit testing* and the related concept of *test-driven development* [4, 5, 6, 7]—in the context of the OpenWorm project. OpenWorm is an international, collaborative open-science project aimed at integrating the world's collective scientific understanding of the *C. elegans* round worm into a single computational model [8]. It is a diverse project incorporating data, simulations, powerful but intuitive user interfaces, and visualization. Since the goal of the project is to simulate an entire organism, the project and its underlying code are necessarily complex. The scope of the project is immense – OpenWorm has over fifty contributors from sixteen countries and projects divided into over forty-five sub-repositories under version control containing a total of hundreds of thousands of lines of code. For a project of this magnitude to remain manageable and sustainable, a thorough testing framework and culture of test-driven development is essential [4, 5, 6, 7]. In Figure 1, we show a diagrammatic overview of the many projects within OpenWorm and the relationship of testing to each of these. For extremely small projects, unit testing simply adds an overhead with little or no return on the time investment. As the project grows in size, however, the gains are quite significant, as the burden on the programmers of maintaining a large project can be substantially reduced.

In the code excerpts below, we will discuss 4 types of tests that are used in the OpenWorm code-base. They are:

- **Verification tests**: These are tests of basic software correctness and are not unique to the scientific nature of the project.
- **Data integrity tests**: These are tests unique to a project which incorporates data. Among other purposes, these tests serve as basic sanity checks verifying, for instance, that each piece of data in the project is associated with a scientific paper and corresponding DOI.
- **Biological integrity tests**: These are tests that verify correspondence with known information about static parameters that characterize *C. Elegans*, for example, the total number of neurons.
- **Model validation tests**: These are tests unique to projects which incorporate dynamic models. Model validation tests (using the Python package `SciUnit`) verify that a given dynamic model (such as the behavior of an ion channel) generates output that is consistent with known behavior from experimental data.

The target audience for this article is computational biologists who have limited experience with large software projects and are looking to incorporate standard industrial practices into their work, or who anticipate involvement with larger projects in either academia or industry. We also hope that the exposition will be accessible to other scientists interested in learning about computational techniques and software engineering. We hope to contribute to raising the quality of biological software by describing some basic concepts of software engineering in the context of a practical research project.

# Unit testing for scientific software

## A simple introduction to unit testing

The basic concept behind software testing is quite simple. Suppose we have a piece of code which takes some number of inputs and produces corresponding outputs. A *unit test*, *verification test*, or simply *test* is a function that compares an input-output pair and returns a boolean value *True* or *False*. A result of *True* indicates that the code is behaving as intended, and a result of *False* indicates that it is not, and consequently, that any program relying on that code cannot be trusted to behave as intended.

Let us take a simple example. Suppose we have a function that takes a list of numbers and then returns them in sorted order, from lowest to highest. Sorting is a classic algorithmic task, and there are many different sorting algorithms with different performance characteristics; while the specific strategies they employ differ wildly, ultimately the result should be the same for any implementation. A unit test for one's sorting algorithm should take as input a list of numbers, feed it to the sorting algorithm, and then check that each element in the output list is less than or equal to the one that comes after it. The unit test would return True if the output list had that property, and False if not.

If one has multiple implementations of a sorting algorithm, then one can use a reliable reference implementation as a testing mechanism for the others. In other words, a test might return *True* if a novel sorting algorithm gives the same result as one widely known to be valid. There are other strategies along these lines. For example, suppose we have an implementation of an algorithm for multiplication called `multiply`. If we have a trusted implementation of an algorithm for addition, we can test that our multiplication algorithm works as expected by checking its behavior against the appropriate number of addition operations, e.g., `multiply(3,5) = 3 + 3 + 3 + 3 + 3`. See Listing 1 for an implementation of this test in Python code.

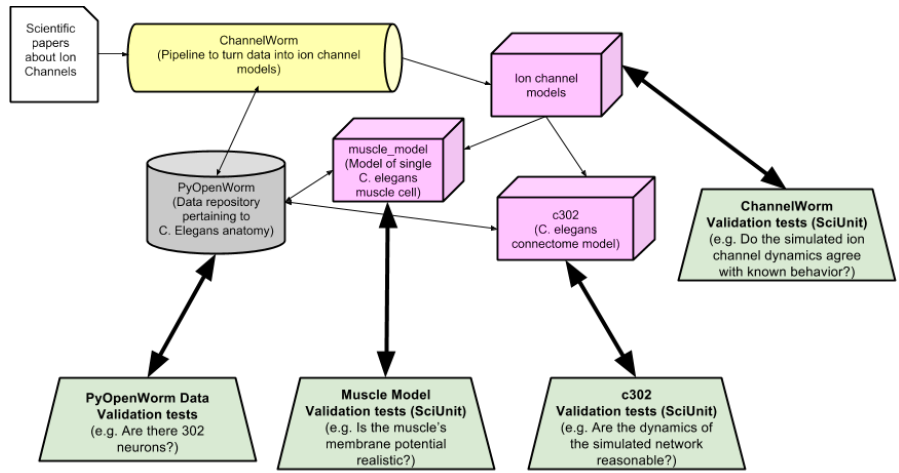

**Figure 1.** Diagram of the some key OpenWorm modules and their corresponding testing frameworks.

In the previous example, our unit test verified the core functionality of the algorithm. We had an algorithm that claimed to sort things, and we wanted to check that it worked as advertised. But there are many other kinds of tests that we might be compelled to write in order to know that our software is working correctly. For instance, what happens if we feed an empty list to our sorting algorithm (this is an example of an *edge case*)? Should it simply return the list, generate an error message, or both? What if a user accidentally gives the algorithm something that is not a list, say for example, an image? What should the error message be in this case? Should there be a single error message to cover all cases, or should the error message be tailored to the specific case at hand? One can easily write unit tests to verify that the correct behavior has been implemented in all of these cases.

The sum total of all of the desired behaviors of an algorithm is called a *specification*, or *spec* for short. For instance, the specification for a sorting algorithm might look like the following:

- When given a list of numbers, return the list sorted from smallest to largest.
- When given a list of strings, return the list sorted in lexicographic order.
- If the input is an empty list, return the empty list and do not generate an error message.
- If the input is not a list, generate the error message "Input should be a list of real numbers or strings".
- If the input is neither a list of strings nor a list of numbers, return the same error message as above.

In Listing 2, we have given a suite of units tests for a sorting algorithm called `mySort` based on this specification. The basic notion demonstrated there in the context of the sorting algorithm extends to any piece of software. In OpenWorm, we make extensive use of unit testing to verify both the functional properties of the system, as well as the validity of the data and models that comprise the simulation. For instance, the two tests given below in Listing 3 check that any worm model has 302 neurons, and that the number of synapses for a given type of neuron is in accordance with its known value from the scientific literature. We will examine the different types of tests in more detail in the next section.

In *test-driven development*, the specification for a piece of software, as well as the corresponding unit tests are written *before coding the software itself* [4, 7]. The argument for test-driven development is that having a well-developed testing framework before beginning the actual process of software development increases the likelihood that bugs will be caught as quickly as possible, and furthermore, that it helps the programmer to clarify their thought processes. In practice, while some tests are written before-hand, others are written in parallel with the rest of code development, or shortly after a piece of code is written but before it is integrated.

We mention here that, in the software community, a distinction is often made between unit tests and *integration*

**Listing 1.** Simple test for the multiplication operation.

```python
def test_multiply():
    """
    Test our multiplication function against the
    standard addition operator
    """
    assert multiply(3, 5) == 3 + 3 + 3 + 3 + 3
```

**Listing 2.** Sample tests for the sorting specification given in the text. The class SortingTest is a container for all of the individual tests that define the specification and can be extended if more tests are added.

```python
import random
import unittest
from my_code import my_sort

"""
Specification:
1) When given a list of numbers,
return the list sorted from smallest to largest.

2) When given a list of strings,
return the list sorted in lexicographic order.

3) If the input is an empty list,
return the empty list and do not generate an error message.

4) If the input is not a list, generate the error message:
``Input should be a list of real numbers or strings''.
"""

class SortingTest(unittest.TestCase):
    """A class implementing tests for a sorting function"""
    def setUp(self):
        self.f = my_sort # The function we will test is mySort

    def test_number_sort(self):
        """Test that numbers sort correctly"""
        sorted_list = range(100000)
        shuffled_list = random.shuffle(range(100000))
        self.assertEqual(self.f(shuffled_list), sorted_list)

    def test_string_sort(self):
        """Test that strings sort correctly"""
        word_file = '/usr/share/dict/words'
        words = open(word_file).read().splitlines()
        sorted_words = words
        shuffled_words = random.shuffle(words)
        self.assertEqual(self.f(shuffled_words), sorted_words)

    def test_empty_list(self):
        """Test that empty list returns empty list"""
        self.assertEqual(self.f([]), [])

    def test_not_list(self):
        """Test that invalid inputs generate correct error message"""
        message = 'Input should be a list of real numbers or strings.'
        self.assertRaisesRegexp(TypeError, message, self.f, 'a')

    def test_mixed_list(self):
        """Test that mixed lists generate appropriate error message"""
        mixed_list = [1, 2, 'a', 'b', 3]
        message = 'Input should be a list of real numbers or strings.'
        self.assertRaisesRegexp(TypeError, message, self.f, mixed_list)
```

*tests* [7]. Strictly speaking, a unit test is a test which is applicable to the smallest, functional unit of code, and which has no external dependencies. On the other hand, tests which verify that different components work together are classified as integration tests; they verify that multiple components are integrated correctly. Some of the tests discussed below would strictly be considered integration tests. For the sake of simplicity, we will not distinguish between unit tests and integration tests in this article, and will refer to both as simply *tests* or *unit tests*. The primary distinction that we make here is instead between ordinary *verification* tests (to verify that code works as intended) and *model validation* tests (to validate a model against experimental data), which we discuss in more depth below.

### Unit testing in OpenWorm

The software that makes up OpenWorm shares common ground with all other pieces of software, whether the sorting algorithm described above, a word processor, or an operating system. As a result, there are a range of unit tests that need to be written to ensure that basic pieces of the software infrastructure function correctly. Many of these tests will not be of any scientific significance; they are simply sanity checks to ensure correct behavior for predictable cases. For instance, there are tests for checking that certain internal functions return the appropriate error messages when given incorrect inputs; there are tests for verifying that databases are loaded correctly; there are tests which check that functions adhere to a specific naming convention which will help automated tools process the code-base.

As a data-driven, scientific research project, however, OpenWorm also makes use of several other categories of tests that do not typically appear in software development. For instance, the `PyOpenWorm` subproject of OpenWorm is a simple API that provides a repository of information about *C. elegans* anatomy (https://github.com/openworm/PyOpenWorm). Given that the aim of OpenWorm is to produce a realistic simulation of the nematode, a carefully curated repository of empirical information is a cornerstone of the project.

In the context of unit testing, there needs to be a category of tests that ensure that a curated datum has been appropriately verified and, furthermore, that its internal representation in the `PyOpenWorm` database is consistent. For example, for each "fact" in `PyOpenWorm`, there needs to be an associated piece of evidence, which serves as a reference. Practically, this evidence consists of a Digital Object Identifier [9], or DOI, which corresponds to a research paper from which the fact was originally retrieved. For this class of tests, we traverse the database of facts and verify that for each fact there is an associated source of evidence, i.e., a DOI. Furthermore, these tests verify that each DOI is valid, and that the URL corresponding to the DOI is accessible. There are also tests to check the internal consistency of the `PyOpenWorm` database, for instance, that neurons with the same name have the same identifier.

Listing 4 gives several excerpts from the `PyOpenWorm` testing framework. It consists of tests to verify the references in the database, i.e., the DOIs which correspond to research papers.

In Listing 5, we give several tests for verifying the contents of the `PyOpenWorm` repository. Since each of the functions below is designed to test properties of `Neuron` objects, they are part of a single class called `NeuronTest`. These tests fall into the category of verification tests, and several of the tests, such as `test_name` and `test_type` simply check that the database is working correctly.

### Model validation with `SciUnit`

Many computational models in biology are compared only informally with the experimental data they aim to explain. In contrast, we formalize data-driven model validation in OpenWorm by incorporating tests to *validate* each dynamical model in the project against experimental data from the literature. As an example, consider a scenario where a developer creates a new model and provides parameter values for a simulation. In addition to running all of the *verification* tests described above, the model and parameter values must be *validated* with respect to established experimental results. In general, each summary output of the model is validated against a corresponding piece of data. One example of a summary model output is the "IV Curve" (i.e. current evoked in response to each of a series of voltage steps) of a given neuronal ion channel. We expect that our model will possess only ion channels which behave similarly to those observed experimentally, i.e. that the model IV Curve matches the experimentally-determined IV curve. If our model's IV curve deviates too greatly from that observed experimentally, the model developers should be alerted and provided with information that will allow them to investigate the source of the discrepancy [10]. This may mean that parameter values must be modified, or in some cases the model itself must be substantially revised. In the case of OpenWorm, the necessary data for validating models is part of the `PyOpenWorm` and `ChannelWorm` subprojects (https://github.com/openworm/ChannelWorm), which are repositories of curated information about *C. elegans* anatomy and ion channels.

Ordinary unit testing frameworks do not readily lend themselves to this kind of model validation. Rather than simply comparing an input-output pair, model validation tests should perform the same procedure that a scientist would perform before submitting a newly hypothesized model for publication. That is, they should generate some kind of summary statistic encoding the deviation between experimental data and model output. For example, in the case of an IV Curve, one might use the area between

**Listing 3.** Excerpts from basic biological integrity tests for worm models. Given the size of the data repositories that OpenWorm relies upon, there are many simple tests such as these for ensuring the correctness of the associated data.

```python
import PyOpenWorm
import unittest

class BiologicalIntegrityTest(unittest.TestCase):
    """
    Tests that read from the database and ensure that basic
    queries have expected results, as a way to ensure data quality.
    """
    def test_correct_neuron_number(self):
        """
        This test verifies that the worm model
        has exactly 302 neurons.
        """
        net = PyOpenWorm.Worm().get_neuron_network()
        self.assertEqual(302, len(set(net.neurons())))

    def test_neuron_syn_degree(self):
        """
        This test verifies that the number of chemical synapses
        associated with a given neuron AVAL is equal to 90.
        """
        aval = PyOpenWorm.Neuron(name='AVAL')
        self.assertEqual(aval.Syn_degree(), 90)
```

**Listing 4.** Verifying data integrity is an integral component of testing in OpenWorm. Below, we give several sample tests to verify the existence of valid DOIs, one technique used to ensure that facts in the `PyOpenWorm` repository are appropriately linked to the research literature.

```python
import _DataTest # our in-house setup/teardown code
from PyOpenWorm import Evidence

class EvidenceQualityTests(_DataTest):
    """A class implementing tests for evidence quality."""
    def test_has_valid_resource(self):
        """Checks if the object has either a valid DOI or URL"""
        ev = Evidence()
        allEvidence = list(ev.load())
        evcheck = []

        """Loop over all evidence fields in the database"""
        for evobj in allEvidence:
            if evobj.doi():
                doi = evobj.doi()
                val = requests.get('http://dx.doi.org/' + doi)
                evcheck.append(val.status_code == 200)

            elif evobj.url():
                url = evobj.url()
                val = requests.get(url)
                evcheck.append(val.status_code == 200)

            else:
                evcheck.append(False)

        self.assertTrue(False not in evcheck)
```

**Listing 5.** An assortment of verification tests from `PyOpenWorm`. These verify that the database behaves as we would expect it to, that properties of certain objects (`Neuron` objects, in this case) are correctly specified, and that the database is not populated with duplicate entries.

```python
import _DataTest # our in-house setup/teardown code
from PyOpenWorm import Neuron

class NeuronTest(_DataTest):
    """
    AVAL, ADAL, AVAR, and PCVL are individual neurons in C. Elegans.
    AB plapaaaap is the lineage name of the ADAL neuron.
    A class implementing tests for Neuron objects.
    """
    def test_same_name_same_id(self):
        """
        Test that two Neuron objects with the same name
        have the same identifier().
        """
        c = Neuron(name='AVAL')
        c1 = Neuron(name='AVAL')
        self.assertEqual(c.identifier(query=True), c1.identifier(query=True))

    def test_type(self):
        """
        Test that a Neuron's retrieved type is identical to
        its type as inserted into the database.
        """
        n = self.neur('AVAL')
        n.type('interneuron')
        n.save()
        self.assertEqual('interneuron', self.neur('AVAL').type.one())

    def test_name(self):
        """
        Test that the name property is set when the neuron
        is initialized with it.
        """
        self.assertEqual('AVAL', self.neur('AVAL').name())
        self.assertEqual('AVAR', self.neur('AVAR').name())

    def test_init_from_lineage_name(self):
        """
        Test that we can retrieve a Neuron from the database
        by its lineage name only.
        """
        c = Neuron(lineageName='AB plapaaaap', name='ADAL')
        c.save()
        c = Neuron(lineageName='AB plapaaaap')
        self.assertEqual(c.name(), 'ADAL')

    def test_neighbor(self):
        """
        Test that a Neuron has a 'neighbors' property, and that the
        correct Neuron is returned when calling the 'neighbor' function.
        """
        n = self.neur('AVAL')
        n.neighbor(self.neur('PVCL'))
        neighbors = list(n.neighbor())
        self.assertIn(self.neur('PVCL'), neighbors)
        n.save()
        self.assertIn(self.neur('PVCL'), list(self.neur('AVAL').neighbor()))
```

the model and data curves as a summary statistic. In the case of OpenWorm, because these models are part of a continuously updated and re-executed simulation, and not simply static equations in a research paper, the model validation process must happen automatically and continuously, alongside other unit tests.

To incorporate model validation tests, we use the Python package SciUnit [11]. While there are some practical differences between writing SciUnit tests and ordinary unit tests, the concepts are quite similar. For example, a SciUnit test can be configured to return *True* if the test passes, i.e. model output and data are in sufficient agreement, and *False* otherwise. Ultimately, a scientific model is just another piece of software—thus it can be validated with respect to a specification. In the case of dynamical models, these specifications come from the scientific literature, and are validated with the same types of tests used before submitting a model for publication. SciUnit simply formalizes this testing procedure in the context of a software development work-flow. In Listing 6, we give an example of SciUnit tests using the neuron-specific helper library NeuronUnit (http://github.com/scidash/neuronunit) for neuron-specific models).

In the preceding example, the statistic is computed within the SciUnit method judge, which is analogous to the self.assert statements used in the ordinary unit tests above. While the ordinary unit test compares the output of a function pair to an accepted reference output, judge compares the output of a model (i.e. simulation data) to accepted reference experimental data. Internally, the judge method invokes other code (not shown) which encodes the test's specification, i.e. what a model must do to pass the test. The output of the test is a numeric score. In order to include SciUnit tests alongside other unit tests in a testing suite, they can be configured to map that numeric score to a boolean value reflecting whether the model/data agreement returned by judge is within an acceptable range.

The output of these model validation tests can also be inspected visually; Figure 2 shows the graphical output of the test workflow in Listing 6, and illustrates for the developers why the test failed (mismatch between current-voltage relationship produced by the model and the one found in the experimental literature). Further details about the output of this test – including the algorithm for computing model/data agreement, and the magnitude of disagreement required to produce a failing score – can also be accessed via attributes and methods of the score object (not shown, but see SciUnit documentation). Consequently, full provenance information about the test is retained.

Some computational science projects use ad-hoc scripts that directly run models and compare their outputs to reference data. This can be adequate in simple cases, but for larger projects, particularly, distributed open-source projects with many contributors, the mixing of implementation and interface carries significant drawbacks [12]. For example, in order to record and store the membrane potential of a model cell–to then compare to reference data–one could determine which functions are needed to run the simulation in a given simulation engine, extract the membrane potential from the resulting files, and then call those functions in a test script. However, this approach has three major flaws. First, it may be difficult for a new contributor or collaborator to understand what is being tested, as the test code is polluted with implementation details of the model that are not universally understood. Second, such a test will not work on any model that does not have the same implementation details, and thus has limited re-usability. Third, any changes to the model implementation will require parallel changes to the corresponding tests. In contrast, by separating tests from implementation, tests can work on any model that implements a well-defined set of capabilities exposed via an interface. SciUnit does this by design, and SciUnit tests interact with models only through an interface of standard methods, for example, those provided by NeuronUnit. It is the responsibility of the model developer to match this interface by referencing standard methods, e.g. run, get_membrane_potential, etc. Ultimately, the separation of implementation from interface leads to greater code clarity, more rapid development, and greater test re-usability.

## Test coverage

The *coverage* of a testing suite is defined as the percentage of functions in a code-base which are being tested. Since there is no rigorous measure of what constitutes an adequate test, precise figures of test coverage should be interpreted with caution. Nonetheless, automated tools which analyze a code-base to determine test coverage can be a valuable resource in suggesting areas of a code-base in need of additional attention. Ideally, test coverage should be as high as possible, indicating that a large fraction of or even the entire code-base has been tested according to the intended specifications.

In PyOpenWorm, we make use several of pre-existing tools in the Python ecosystem for calculating test coverage of the Python code-base, specifically, the aptly-named Coverage package [13], as well as a GitHub extension dedicated to tracking the coverage of such projects known as Coveralls [14]. We adopted these tools in an effort to track which parts of the code-base need additional tests, and to give further backing to the test-driven culture of the project. PyOpenWorm currently has a test coverage of roughly 73%. If a contributor to PyOpenWorm introduces some new code to the project but does not add tests for it, the contributor will see that test coverage has been reduced. By making changes in test coverage explicit, for example, with a badge on the project's homepage, it is easier to track the impact of a growing code-base.

**Listing 6.** Excerpt from a `SciUnit` test in `ChannelWorm`, a repository of information about ion channels. The test listed here verifies that a given ion channel has the correct current / voltage behavior. In terms of the informal classification of tests given above, this test falls under the category of model validation tests.

```
import os, sys
import numpy as np
import quantities as pq
from neuronunit.tests.channel import IVCurvePeakTest
from neuronunit.models.channel import ChannelModel
from channelworm.ion_channel.models import GraphData

# Instantiate the model; CW_HOME is the location of the ChannelWorm repo
ch_model_name = 'EGL-19.channel'
channel_id = 'ca_boyle'
ch_file_path = os.path.join(CW_HOME, 'models', '%s.nml' % ch_model_name)
model = ChannelModel(ch_file_path, channel_index=0, name=ch_model_name)

# Get the experiment data and instantiate the test
doi = '10.1083/jcb.200203055'
fig = '2B'
sample_data = GraphData.objects.get(
        graph__experiment__reference__doi=doi,
        graph__figure_ref_address=fig
    )

# Current density in A/F and membrane potential in mV.
obs = zip(*sample_data.asarray())
observation = {'i/C':obs[1]*pq.A/pq.F, 'v':obs[0]*pq.mV}

# Use these observations to instantiate a quantitative test of the peak
# current (I) in response to a series of voltage pulses (V) delivered
# to the channel.
test = IVCurvePeakTest(observation)

# Judge the model output against the experimental data.
# Score will reflect a measure of agreement between I/V curves.
score = test.judge(model)
score.plot()
```

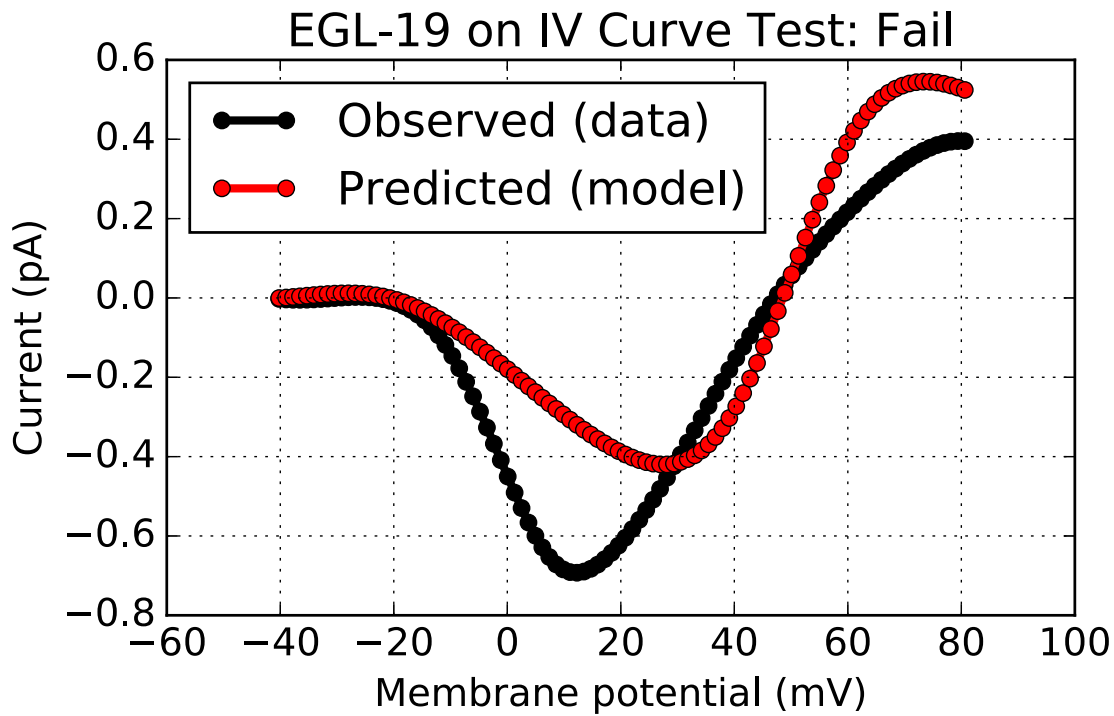

**Figure 2.** Graphical output from Listing 6, showing a failed test which alerts developers to an inconsistency between model and data.

### Continuous integration

Modern software is often written using a process of *continuous integration* or CI [15, 16], whereby the contributions of developers are integrated into a shared repository multiple times a day by an automated system. Typically, the output of a testing suite will determine whether or not the new contributions of a developer can be immediately integrated, or whether changes are required to avoid *regression*, i.e. failing unit tests that were passing before the new contribution.

The benefits of continuous integration include early detection of bugs, eliminating development bottle-necks close to the release date (in the case of commercial software), and the regular availability of usable versions of the software. The process of continuous integration also encourages shifts in how developers think about structuring their code, and encourages regular, modular contributions, rather than massive, monolithic changes that can be difficult to debug.

The entire OpenWorm project, including the `PyOpenWorm` and `ChannelWorm` modules make use of continuous integration (see Fig. 3), taking advantage of a free service called Travis-CI that tests changes to the code-base as they are pushed to the collaborative software development portal GitHub [17]. With each change, the entire project is built from scratch on a machine in the cloud, and the entire test suite is run. A build that passes all tests is a "passing build", and the changes introduced will not break any functionality that is being tested. Because the entire project is built from scratch with each change to the code-base, the dependencies required to achieve this build must be made explicit. This ensures that there is a clear roadmap to the installation of dependencies required to run the project successfully – no hidden assumptions about pre-existing libraries can be made.

### Skipped tests and expected failures

Suppose we have rigorously employed a process of test-driven development. Starting with a carefully designed specification, we have written a test suite for a broad range of functionality, and are using a continuous integration system to incorporate the ongoing contributions of developers on a regular basis.

In this scenario, given that we have written a test suite prior to the development of the software, our CI system will reject all of our initial contributions because most tests fail, simply because the code that would pass the tests has not been written yet! To address precisely this scenario, many testing frameworks allow tests to be annotated as *expected failures* or simply to skip a given test entirely. The ability to mark tests as expected failures allows developers to incrementally enable tests, and furthermore, draws attention to missing functionality. Consequently, the fraction of tests passed becomes a benchmark for progress towards an explicit development goal, that goal being encoded by the set of all tests that have been written.

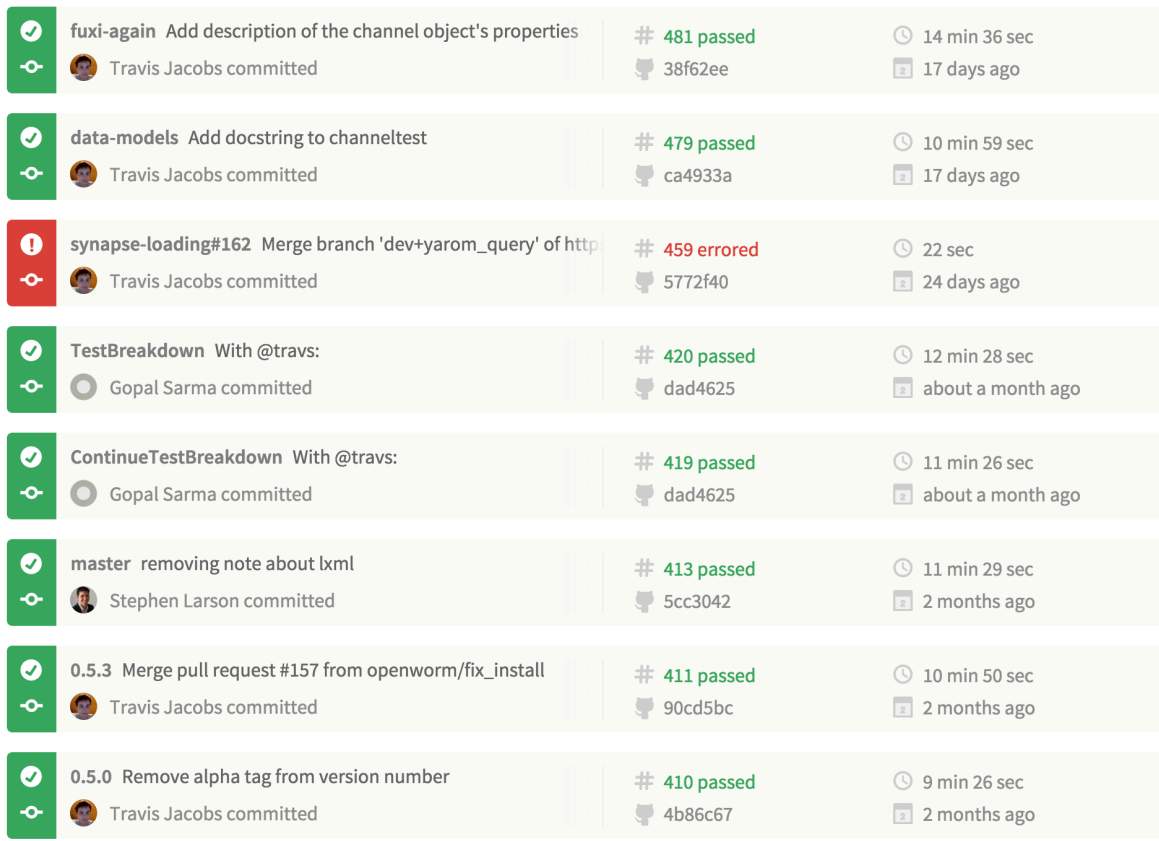

**Figure 3.** Sample output from the OpenWorm continuous integration dashboard. Each row corresponds to a single set of contributions, known as a *commit*, submitted by a given developer. A commit is assigned a *build number*, which is given in the second column, and the result of the build process is indicated by the color of the corresponding row. If any of the unit tests fail, the build will be marked as failed, and the code contributions will be rejected. The developer is then responsible for identifying and fixing the corresponding bugs, and resubmitting their contributions to the code repository.

The OpenWorm code-base makes extensive use of skipped tests and expected failures as a core part of the culture of test-driven development. In `PyOpenWorm`, for example, data integrity tests are often added in advance of the data itself being incorporated to the database. These tests provide a critical safety net as new information is curated from the scientific literature. Prior to the curation of this information, the tests are simply skipped. Once the information is curated, the tests are run, and indicate whether the information is usable by the project.

### Frivolous tests and overly specific tests

Tests are typically sufficiently straightforward to write that it is easy to proliferate a testing suite with a large number of unnecessary tests. Often, these tests will be completely frivolous and cause no harm, beyond causing a testing suite to take much longer than necessary to run. However, tests which are overly specific can actually hinder the process of development. If there are tests which are too specific and constrain internal behavior that is not meant to be static, a developer's contributions may be unnecessarily rejected during the process of continuous integration.

## Conclusions

Our aim in this article is to give an overview of some basic development practices from industrial software engineering that are of particular relevance to biological software. As a summary, we list here the types of tests used in OpenWorm. This list is simply an informal classification, and not a definitive taxonomy:

**Verification tests (the usual suspects)** These are tests common to all pieces of software and are not particu-
Page 11 of 13

larly relevant to the biological nature of the project. For instance, tests that verify that error handling is implemented correctly, that databases are accessed correctly, or that performing certain numerical operations produces results within an acceptable range.

**Data integrity tests**   These are tests unique to a project that incorporates curated data. In the case of OpenWorm, these tests check (among other things) that every biological fact in the `PyOpenWorm` repository has an associated piece of experimental evidence, typically corresponding to a DOI, and that each of these DOIs is valid.

**Biological integrity tests**   These tests verify that data tokens in the `PyOpenWorm` repository correspond to known information about *C. Elegans*. In contrast to the model validation tests described below, biological integrity tests typically only check static information / parameters.

**Model validation tests**   These are tests specific to a project that incorporates scientific models. Model validation tests allow us to check that specific models, such as the behavior of ion channels, correspond to known behavior from the scientific literature. In effect, they extend the notion of unit testing to compare summary data and model output according to some summary statistic. In OpenWorm, the Python package `SciUnit` and derivative packages like `NeuronUnit` are used for writing tests that check the validity of scientific models against accepted data.

It should be clear from the above discussion and corresponding code examples that unit tests are fundamentally quite simple objects. Their behavior is no more than to compare input-output pairs, or in the case of `SciUnit` tests, that a given model's output corresponds to a known reference from the scientific literature. The sophistication of testing frameworks is generally quite minimal when compared to the software itself being tested. While ad-hoc test scripts may be sufficient for small projects, for large projects with many contributors, a systematic approach to unit testing can result in significant efficiency gains and ease the burden of long-term code maintenance. In the context of *continuous integration*, whereby a piece of software is built in an ongoing cycle as developers make changes and additions to the code-base, unit testing provides a valuable safety net that can prevent flawed code from prematurely being integrated.

However, in spite of the conceptual simplicity and potential pitfalls of testing, its importance cannot be overstated. Writing tests requires careful thought and planning and some knowledge of the code-base being tested. Testing from a specification alone can result in inadequate testing, but tests which are too specific to the code-base can result in unnecessary roadblocks for developers.

Rather than being thought of as a sophisticated set of technical tools, unit testing should be viewed as a cultural practice for ensuring the reliability of complex software. Perhaps a useful analogy is the powerful impact that checklists have had in clinical medicine, aviation, construction, and many other industries [18, 19, 20]. Unit tests are sanity checks at a minimum, and can potentially guide the scientific development of models when used in conjunction with experimental data. In order to reap their benefit, their existence and maintenance needs to be valued by all of the participants of the research and software development process. Finally, in order for this culture to be created, test-driven development should not be a heavy-handed imposition on the developers. Otherwise, it will be incorrectly perceived as a bureaucratic hurdle, rather than the valuable safety-net that it is.

### Author contributions
GPS, TWJ, RCG, and SDL wrote the manuscript. All authors contributed to the unit testing framework in the Open Worm project.

### Grant information
This work was funded in part by NIMH (R01MH106674), and NIBIB (R01EB021711 and R01EB014640).

### Acknowledgements
We would like to thank supporters of Open-Worm, including NeuroLinx.org, the backers of the 2014 OpenWorm Kickstarter campaign (http://www.openworm.org/supporters.html), Google Summer of Code 2015, and the International Neuroinformatics Coordinating Facility. We would also like to thank the scientific and code contributors to OpenWorm (http://www.openworm.org/people.html), and Shreejoy Tripathy for careful reading of the manuscript.